\documentstyle[times,pramana,epsf,floats]{ias}

\newcommand{\AmS}{{\protect\the\textfont2
  A\kern-.1667em\lower.5ex\hbox{M}\kern-.125emS}}

\begin{document}
\newcommand{\be}{\begin{equation}}
\newcommand{\ee}{\end{equation}}
\newcommand{\bea}{\begin{eqnarray}}
\newcommand{\eea}{\end{eqnarray}}
\newcommand{\ba}{\begin{array}}
\newcommand{\ea}{\end{array}}
\newcommand {\Dslash}{D\!\!\!/}

\mark{{Superdense Matter}{Thomas Schafer}}
\title{Superdense Matter}

\author{Thomas Sch\"afer}
\address{  Department of Physics and Astronomy, 
     State University of New York, 
     Stony Brook, NY 11794-3800 and
     Riken-BNL Research Center, Brookhaven National 
     Laboratory, Upton, NY 11973 }
\abstract{We review recent work on the phase structure of QCD 
at very high baryon density. We introduce the phenomenon of
color superconductivity and discuss the use of weak coupling
methods. We study the phase structure as a function of the 
number of flavors and their masses. We also introduce 
effective theories that describe low energy excitations 
at high baryon density. Finally, we study the possibility 
of kaon condensation at very large baryon density.}

\maketitle

\section{Color Superconductivity}
\label{sec_intro}

  In the interior of compact stars matter is compressed
to densities several times larger than the density of 
ordinary matter. Unlike the situation in relativistic
heavy ion collisions, these conditions are maintained 
for essentially infinite periods of time and the material 
is quite cold. At low density quarks are confined, chiral 
symmetry is broken, and baryonic matter is described in 
terms of neutrons and protons as well as their excitations.
At very large density, on the other hand, we expect that 
baryonic matter is described more effectively in terms
of quarks rather than hadrons. As we shall see, these
quarks can form new condensates and the phase structure
of dense quark matter is quite rich. 

  At very high density the natural degrees of freedom are
quark excitations and holes in the vicinity of the Fermi
surface. Since the Fermi momentum is large, asymptotic freedom
implies that the interaction between quasi-particles is weak.
In QCD, because of the presence of unscreened long range gauge 
forces, this is not quite true. Nevertheless, we believe 
that this fact does not essentially modify the argument. 
We know from the theory of superconductivity the Fermi 
surface is unstable in the presence of even an arbitrarily 
weak attractive interaction. At very large density, the 
attraction is provided by one-gluon exchange between quarks 
in a color anti-symmetric $\bar 3$ state. High density quark 
matter is therefore expected to behave as a color superconductor 
\cite{Frau_78,Barrois:1977xd,Bar_79,Bailin:1984bm}.

  Color superconductivity is described by a pair condensate
of the form
\be
\label{csc}
\phi = \langle \psi^TC\Gamma_D\lambda_C\tau_F\psi\rangle.
\ee
Here, $C$ is the charge conjugation matrix, and $\Gamma_D,
\lambda_C,\tau_F$ are Dirac, color, and flavor matrices. 
Except in the case of only two colors, the order parameter
cannot be a color singlet. Color superconductivity is 
therefore characterized by the breakdown of color gauge 
invariance. As usual, this statement has to be interpreted 
with care because local gauge invariance cannot really be 
broken. Nevertheless, we can study gauge invariant 
consequences of a quark pair condensate, in particular
the formation of a gap in the excitation spectrum.

 In addition to that, color superconductivity can lead to 
the breakdown of global symmetries. We shall see that in 
some cases there is a gauge invariant order parameter 
for the $U(1)$ of baryon number. This corresponds to true 
superfluidity and the appearance of a massless phonon. 
We shall also find that for $N_f>2$ color superconductivity 
leads to chiral symmetry breaking. Finally, if the effects
of finite quark masses are taken into account we find
additional forms of long-range order.

\section{Phase Structure in Weak Coupling}
\label{sec_phases}
\subsection{QCD with two flavors}
\label{sec_nf2}

  In this section we shall discuss how to use weak coupling
methods in order to explore the phases of dense quark matter.
We begin with what is usually considered to be the simplest 
case, quark matter with two degenerate flavors, up and down. 
Renormalization group arguments suggest 
\cite{Evans:1999ek,Schafer:1999na}, and explicit
calculations show \cite{Brown:1999yd,Schafer:2000tw}, that
whenever possible quark pairs condense in an $s$-wave. This
means that the spin wave function of the pair is anti-symmetric. 
Since the color wave function is also anti-symmetric, the Pauli
principle requires the flavor wave function to be anti-symmetric
too. This essentially determines the structure of the order
parameter \cite{Alford:1998zt,Rapp:1998zu}
\be
\phi^a  = \langle \epsilon^{abc}\psi^b C\gamma_5 \tau_2\psi^c
 \rangle.
\ee
This order parameter breaks the color $SU(3)\to SU(2)$ and
leads to a gap for up and down quarks with two out of the 
three colors. Chiral and isospin symmetry remain unbroken. 

  We can calculate the magnitude of the gap and the 
condensation energy using weak coupling methods. In weak
coupling the gap is determined by ladder diagrams with 
the one gluon exchange interaction. These diagrams can 
be summed using the gap equation 
\cite{Son:1999uk,Schafer:1999jg,Pisarski:2000tv,Hong:2000fh,Brown:1999aq}
\bea
\label{eliash}
\Delta(p_0) &=& \frac{g^2}{12\pi^2} \int dq_0\int d\cos\theta\,
 \left(\frac{\frac{3}{2}-\frac{1}{2}\cos\theta}
            {1-\cos\theta+G/(2\mu^2)}\right. \\
 & & \hspace{3cm}\left.    +\frac{\frac{1}{2}+\frac{1}{2}\cos\theta}
            {1-\cos\theta+F/(2\mu^2)} \right)
 \frac{\Delta(q_0)}{\sqrt{q_0^2+\Delta(q_0)^2}}. \nonumber
\eea
Here, $\Delta(p_0)$ is the frequency dependent gap, $g$ is the 
QCD coupling constant and $G$ and $F$ are the self energies of
magnetic and electric gluons. This gap equation is very similar
to the BCS gap equations that describe nuclear superfluids. 
The main difference is that because the gluon is massless, 
the gap equation contains a collinear $\cos\theta\sim 1$ 
divergence. In a dense medium the collinear divergence is 
regularized by the gluon self energy. For $\vec{q}\to 0$ 
and to leading order in perturbation theory we have
\be
 F = 2m^2, \hspace{1cm}
 G = \frac{\pi}{2}m^2\frac{q_0}{|\vec{q}|},
\ee
with $m^2=N_fg^2\mu^2/(4\pi^2)$. In the electric part,
$m_D^2=2m^2$ is the familiar Debye screening mass. In the 
magnetic part, there is no screening of static modes, 
but non-static modes are modes are dynamically screened
due to Landau damping.

 We can now perform the angular integral and find
\be
\label{eliash_mel}
\Delta(p_0) = \frac{g^2}{18\pi^2} \int dq_0
 \log\left(\frac{b\mu}{|p_0-q_0|}\right)
    \frac{\Delta(q_0)}{\sqrt{q_0^2+\Delta(q_0)^2}},
\ee
with $b=256\pi^4(2/N_f)^{5/2}g^{-5}$. We can now see why 
it was important to keep the frequency dependence of the 
gap. Because the collinear divergence is regulated by
dynamic screening, the gap equation depends on $p_0$
even if the frequency is small. We can also see that
the gap scales as $\exp(-c/g)$. The collinear divergence 
leads to a gap equation with a double-log behavior. 
Qualitatively
\be
\label{dlog}
 1 \sim \frac{g^2}{18\pi^2}
 \left[\log\left(\frac{\mu}{\Delta}\right)\right]^2,
\ee
from which we conclude that $\Delta\sim\exp(-c/g)$. 
The approximation (\ref{dlog}) is not sufficiently
accurate to determine the correct value of the 
constant $c$. A more detailed analysis shows that
the gap on the Fermi surface is given by
\be
\label{gap_oge}
\Delta_0 \simeq 512\pi^4(2/N_f)^{5/2}\mu g^{-5}
   \exp\left(-\frac{3\pi^2}{\sqrt{2}g}\right).
\ee
We should emphasize that, strictly speaking, this result
contains only an estimate of the pre-exponent. It was 
recently argued that wave function renormalization
and quasi-particle damping give $O(1)$ contributions
to the pre-exponent which substantially reduce the 
gap \cite{Brown:1999aq,Wang:2001aq}.

 For chemical potentials $\mu<1$ GeV, the coupling 
constant is not small and the applicability of perturbation
theory is in doubt. If we ignore this problem and extrapolate
the perturbative calculation to densities $\rho\simeq 5\rho_0$
we find gaps $\Delta\simeq 100$ MeV. This result may indeed 
be more reliable than the calculation on which it is based.
In particular, we note that similar results have been obtained
using realistic interactions which reproduce the chiral 
condensate at zero baryon density \cite{Alford:1998zt,Rapp:1998zu}.

\subsection{QCD with three flavors: Color-Flavor-Locking}
\label{sec_cfl} 

 If quark matter is formed at densities several times 
nuclear matter density we expect the quark chemical 
potential to be larger than the strange quark mass.
We therefore have to determine the structure of the 
superfluid order parameter for three quark flavors.
We begin with the idealized situation of three 
degenerate flavors. From the arguments given in the 
last section we expect the order parameter to be 
color and flavor anti-symmetric matrix of the form
\be
\label{order}
  \phi^{ab}_{ij}=
  \langle \psi^a_i C\gamma_5\psi^b_j\rangle.
\ee
In order to determine the precise structure of this
matrix we have to extremize grand canonical potential.
We find \cite{Schafer:1999fe,Evans:1999at}
\be
\label{cfl}
\Delta^{ab}_{ij} = 
 \Delta (\delta_i^a\delta_j^b-\delta_i^b\delta_j^a),
\ee
which describes the color-flavor locked phase proposed in 
\cite{Alford:1999mk}. Both color and flavor symmetry are completely 
broken. There are eight combinations of color and flavor symmetries 
that generate unbroken global symmetries. The symmetry breaking
pattern is 
\be
\label{sym_3}
SU(3)_L\times SU(3)_R\times U(1)_V\to SU(3)_V .
\ee
This is exactly the same symmetry breaking that QCD exhibits 
at low density. The spectrum of excitations in the color-flavor-locked 
(CFL) phase also looks remarkably like the spectrum of QCD at low 
density \cite{Schafer:1999ef}. The excitations can be classified 
according to their quantum numbers under the unbroken $SU(3)$, and 
by their electric charge. The modified charge operator that generates 
a true symmetry of the CFL phase is given by a linear combination 
of the original charge operator $Q_{em}$ and the color hypercharge 
operator $Q={\rm diag}(-2/3,-2/3,1/3)$. Also, baryon number is only 
broken modulo 2/3, which means that one can still distinguish baryons 
from mesons. We find that the CFL phase contains an octet of Goldstone 
bosons associated with chiral symmetry breaking, an octet of vector 
mesons, an octet and a singlet of baryons, and a singlet Goldstone 
boson related to superfluidity. All of these states have integer 
charges.  

  With the exception of the $U(1)$ Goldstone boson, these states
exactly match the quantum numbers of the lowest lying multiplets
in QCD at low density. In addition to that, the presence of the 
$U(1)$ Goldstone boson can also be understood. The $U(1)$ order
parameter is $\langle (uds)(uds)\rangle$. This order parameter
has the quantum numbers of a $0^+$ $\Lambda\Lambda$ pair condensate.
In $N_f=3$ QCD, this is the most symmetric two nucleon channel, 
and a very likely candidate for superfluidity in nuclear matter
at low to moderate density. We conclude that in QCD with three
degenerate light flavors, there is no fundamental difference 
between the high and low density phases. This implies that a
low density hyper-nuclear phase and the high density quark phase
might be continuously connected, without an intervening phase
transition. 

\subsection{Other phases}
\label{sec_csl}
 
 Color-flavor locking can be generalized to QCD with more than
three flavors \cite{Schafer:1999fe}. Chiral symmetry is broken
for all $N_f\ge 3$, but only in the case $N_f=3$ do we find
the $T=\mu=0$ pattern of chiral symmetry breaking, $SU(N_f)_L
\times SU(N_f)_R\to SU(N_f)_V$. 

 The case $N_f=1$ is special \cite{Schafer:2000tw}. In this
case the order parameter is flavor-symmetric and the Cooper
pairs carry non-zero angular momentum. For $N_c=3$ the spin
direction can become aligned with the color orientation of
the Cooper pair. In the color-spin locked phase color and 
rotational invariance are broken, but a diagonal $SO(3)$ survives. 

 If the number of colors is very large, $N_c\to\infty$,
color superconductivity is suppressed and the ground state 
is very likely a chiral density wave \cite{Deryagin:1992}.
This state is analogous to spin density waves in condensed
matter physics. If the baryon density is large, the transition 
from color superconductivity to chiral density waves 
requires very large values of $N_c$ on the order of
$N_c>1000$ \cite{Shuster:1999}.

\section{The Role of the Strange Quark Mass} 
\label{sec_unlock}

    At baryon densities relevant to astrophysical objects dis\-tor\-tions 
of the pure CFL state due to non-zero quark masses cannot be neglected 
\cite{Alford:1999pa,Schafer:1999pb,Alford:2000ze,Schafer:2000ew,Bedaque:2001je,Kaplan:2001qk}.
This problem can be studied using the effective chiral theory of the 
CFL phase \cite{Casalbuoni:1999wu} (CFL$\chi$Th). This theory determines 
both the ground state and the spectrum of excitations with energies below 
the gap in the CFL phase. Using the effective theory allows us to perform 
systematic calculations order by order in the quark mass. 

\subsection{CFL Chiral Theory (CFL$\chi$Th)}
\label{sec_CFLchi}

 For excitation energies smaller than the gap the only 
relevant degrees of freedom are the Goldstone modes 
associated with the breaking of chiral symmetry and
baryon number. The interaction of the Goldstone modes
is described by an effective Lagrangian of the 
form \cite{Casalbuoni:1999wu}
\bea
\label{l_cheft}
{\cal L}_{eff} &=& \frac{f_\pi^2}{4} {\rm Tr}\left[
 \nabla_0\Sigma\nabla_0\Sigma^\dagger - v_\pi^2
 \partial_i\Sigma\partial_i\Sigma^\dagger \right] 
 +\Big[ C {\rm Tr}(M\Sigma^\dagger) + h.c. \Big] \\ 
 & & \hspace*{-1cm}\mbox{} 
     +\Big[ A_1{\rm Tr}(M\Sigma^\dagger)
                        {\rm Tr} (M\Sigma^\dagger) 
     + A_2{\rm Tr}(M\Sigma^\dagger M\Sigma^\dagger)   
     + A_3{\rm Tr}(M\Sigma^\dagger){\rm Tr} (M^\dagger\Sigma)
         + h.c. \Big]+\ldots . 
 \nonumber 
\eea
Here $\Sigma=\exp(i\phi^a\lambda^a/f_\pi)$ is the chiral field,
$f_\pi$ is the pion decay constant and $M$ is a complex mass
matrix. The chiral field and the mass matrix transform as
$\Sigma\to L\Sigma R^\dagger$ and  $M\to LMR^\dagger$ under 
chiral transformations $(L,R)\in SU(3)_L\times SU(3)_R$.
We have suppressed the singlet fields associated 
with the breaking of the exact $U(1)_V$ and approximate $U(1)_A$ 
symmetries. The theory (\ref{l_cheft}) looks superficially like 
ordinary chiral perturbation theory. There are, however, some 
important differences. Lorentz invariance is broken and Goldstone 
modes move with the velocity $v_\pi<c$. The chiral expansion has 
the structure
\be
{\cal L}\sim f_\pi^2\Delta^2 \left(\frac{\partial_0}{\Delta}\right)^k
 \left(\frac{\vec{\partial}}{\Delta}\right)^l
  \big(\Sigma\big)^m\big(\Sigma^\dagger\big)^n.
\ee
Loop graphs are suppressed by powers of $\partial/(4\pi f_\pi)$.
We shall see that the pion decay constant scales as $f_\pi\sim p_F$.
As a result higher order derivative interactions are parametrically
more important than loop diagrams with the leading order vertices. 

  Further differences as compared to chiral perturbation theory 
in vacuum appear when the expansion in the quark mass is considered. 
The CFL phase has an approximate $(Z_2)_A$ symmetry under which 
$M\to -M$ and $\Sigma\to \Sigma$. This symmetry implies that the 
coefficients of mass terms that contain odd powers of $M$ are 
small. The $(Z_2)_A$ symmetry is explicitly broken by instantons.
The coefficient $C$ can be determined from a weak coupling 
instanton calculation and $C\sim (\Lambda_{QCD}/p_F)^{8}$
\cite{Schafer:1999fe,Manuel:2000wm}. BCS calculations show
that the CFL phase undergoes a phase transition to a less
symmetric phase when $m^2/(2p_F)\sim \Delta$ 
\cite{Alford:1999pa,Schafer:1999pb}. This suggests that the 
expansion parameter in the chiral expansion is $M^2/(p_F\Delta)$.
We shall see that this is indeed the case. However, the 
coefficients $A_{i}$ of the quadratic terms in $M$ 
turn out to be anomalously small
\be
 A_i M^2 \sim \Delta^2 M^2 \sim f_\pi^2\Delta^2 
\left(\frac{M^2}{p_F^2}\right), 
\ee
compared to the naive estimate $A_iM^2\sim f_\pi^2\Delta^2
[M^2/(p_F\Delta)]$.

 The pion decay constant $f_\pi$ and the coefficients $A_i$ can 
be determined using matching techniques. Matching expresses the 
requirement that Green functions in the effective chiral theory 
and the underlying microscopic theory, QCD, agree. The pion decay 
constant is most easily determined by coupling $SU(N_f)_{L,R}$
gauge fields $W_{L,R}$ to the left and right flavor currents. 
As usual, this amounts to replacing ordinary derivatives by 
covariant derivatives. The time component of the covariant 
derivative is given by $\nabla_0\Sigma=\partial_0 \Sigma+
iW_L\Sigma-i\Sigma W_R$ where we have suppressed the vector 
index of the gauge fields. In the CFL vacuum $\Sigma=1$ the 
axial gauge field $W_L-W_R$ acquires a mass by the Higgs 
mechanism. From (\ref{l_cheft}) we get
\be
\label{wm2}
{\cal L} = \frac{f_\pi^2}{4} \, \frac{1}{2} (W_L-W_R)^2.
\ee
The coefficients $A_{1,2,3}$ can be determined by computing
the shift in the vacuum energy due to non-zero quark masses
in both the chiral theory and the microscopic theory. In the 
chiral theory we have 
\be 
\Delta{\cal E}=  
 -\Big[ A_1\left({\rm Tr}(M)\right)^2
      + A_2{\rm Tr}(M^2) + A_3{\rm Tr}(M){\rm Tr} (M^\dagger)
         + h.c. \Big].
\ee
We note that different $O(M^2)$ mass terms produce distinct 
contributions to the vacuum energy. This means that the 
coefficients $A_i$ can be reconstructed uniquely from the 
vacuum energy.

\subsection{High Density Effective Theory (HDET)}
\label{sec_hdet}

 In this section we shall determine the mass of the gauge field
and the shift in the vacuum energy in the CFL phase of QCD
at large baryon density. This is possible because asymptotic
freedom guarantees that the effective coupling is weak. The 
QCD Lagrangian in the presence of a chemical potential
is given by
\be
\label{qcd}
 {\cal L} = \bar\psi \left( i\Dslash +\mu\gamma_0 \right)\psi
 -\bar\psi_L M\psi_R - \bar\psi_R M^\dagger \psi_L 
 -\frac{1}{4}G^a_{\mu\nu}G^a_{\mu\nu},
\ee
where $D_\mu=\partial_\mu+igA_\mu$ is the covariant derivative,
$M$ is the mass matrix and $\mu$ is the baryon chemical 
potential. If the baryon density is very large perturbative QCD 
calculations can be further simplified. The main observation 
is that the relevant degrees of freedom are particle and hole 
excitations in the vicinity of the Fermi surface. We shall describe 
these excitations in terms of the field $\psi_+(\vec{v}_F,x)$, where 
$\vec{v}_F$ is the Fermi velocity. At tree level, the quark field 
$\psi$ can be decomposed as $\psi=\psi_++\psi_-$ where $\psi_\pm
=\frac{1}{2}(1\pm\vec{\alpha}\cdot\hat{v}_F)\psi$. To leading 
order in $1/p_F$ we can eliminate the field $\psi_-$ using 
its equation of motion. For $\psi_{-,L}$ we find
\be
\psi_{-,L} = \frac{1}{2p_F}
  \left( i\vec{\alpha_\perp}\cdot\vec{D}\psi_{+,L}
         + \gamma_0 M \psi_{+,R}\right).
\ee
There is a similar equation for $\psi_{-,R}$.
The longitudinal and transverse components of $\gamma_\mu$ 
are defined by $(\gamma_0,\vec{\gamma})_{\|}=(\gamma_0,\vec{v}
(\vec{\gamma}\cdot\vec{v}))$ and $(\gamma_\mu)_\perp = 
\gamma_\mu-(\gamma_\mu)_{\|}$. To leading order in $1/p_F$ 
the lagrangian for the $\psi_+$ field is given by
\cite{Hong:2000tn,Hong:2000ru,Beane:2000ms}
\bea
\label{hdet}
{\cal L} &=& 
 \psi_{L+}^\dagger (iv\cdot D) \psi_{L+}
  - \frac{ \Delta}{2}\left(\psi_{L+}^{ai} C \psi_{L+}^{bj}
 \left(\delta_{ai}\delta_{bj}-
           \delta_{aj}\delta_{bi} \right) 
           + {\rm h.c.} \right) \nonumber \\ 
& & \hspace{0.5cm}\mbox{}
  - \frac{1}{2p_F} \psi_{L+}^\dagger \left(  (\Dslash_\perp)^2 
  + MM^\dagger \right)  \psi_{L+}  
  + \left( R\leftrightarrow L, M\leftrightarrow M^\dagger \right)  
  + \ldots ,
\eea
with $v_\mu=(1,\vec{v})$ and $i,j,\ldots$ and $a,b,\ldots$ denote 
flavor and color indices.  In order to perform perturbative 
calculations in the superconducting phase we have added a tree 
level gap term $\psi^{ai}_{L,R} C\Delta_{ai,bj} \psi^{bj}_{L,R}$. 
In the CFL phase this term has the structure $\Delta_{ai,bj}
=\Delta(\delta_{ai}\delta_{bj}-\delta_{aj}\delta_{bi})$. The 
magnitude of the gap $\Delta$ is determined order by order in
perturbation theory from the requirement that the thermodynamic
potential is stationary with respect to $\Delta$. With the gap 
term included the perturbative expansion is well defined. There
are no additional infra-red divergences. In particular, there
is no need to include additional gap parameters at higher order
in $1/p_F$, such as the anti-particle gap or modifications of the 
particle gap due to non-zero quark masses \cite{Schafer:2001za}.

\begin{figure}[t]
\epsfxsize=12.5cm
\epsffile{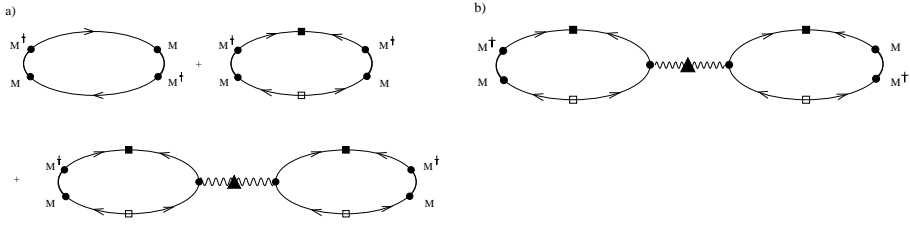}
\vspace*{0.5cm}
\caption{\label{fig_m4}
This figure shows the contribution to the vacuum energy from
the effective chemical potential terms in the high density 
effective theory. Fig.~a) shows the diagrams that are matched 
against the $(MM^\dagger)^2$ term. Fig.~b) shows the diagram
that is matched against the $MM^\dagger \Sigma M^\dagger M
\Sigma$ term in the chiral theory.}
\end{figure}

  The screening mass of the flavor gauge fields $W_{L,R}$ 
can be determined by computing the corresponding polarization 
function in the limit $q_0=0$, $\vec{q}\to 0$. We find 
$\Pi^{LL}_{00}=\Pi^{RR}_{00}=-\Pi^{LR}_{00}=m_D^2/4$ with 
$m_D^2=(21-8\log(2))p_F^2/(36\pi^2)$. Matching this result
against equ.~(\ref{wm2}) we get \cite{Son:1999cm} 
\be
f_\pi^2 = \frac{21-8\log(2)}{18} 
  \left(\frac{p_F^2}{2\pi^2} \right) .
\ee 
Our next task is to compute the mass dependence of the vacuum 
energy. To leading order in $1/p_F$ there is only one operator 
in the high density effective theory
\be 
\label{kin}
{\cal L} = -\frac{1}{2p_F} \left( \psi_{L+}^\dagger MM^\dagger \psi_{L+}
 + \psi_{R+}^\dagger M^\dagger M\psi_{R+} \right).
\ee
This term arises from expanding the kinetic energy of a massive
fermion around $p=p_F$. We note that $MM^\dagger/(2p_F)$ and
$M^\dagger M/(2p_F)$ act as effective chemical potentials 
for left and right-handed fermions, respectively. Indeed, to 
leading order in the $1/p_F$ expansion, the Lagrangian (\ref{hdet}) 
is invariant under a time dependent flavor symmetry $\psi_{L} 
\to L(t)\psi_{L}$, $\psi_{R}\to R(t)\psi_{R}$ where $X_L=
MM^\dagger/(2p_F)$ and $X_R=M^\dagger M/(2p_F)$ transform
as left and right-handed flavor gauge fields. If we impose 
this approximate gauge symmetry on the CFL chiral theory we 
have to include the effective chemical potentials 
$X_{L,R}$ in the covariant derivative of the chiral 
field \cite{Bedaque:2001je}, 
\be
\label{mueff}
 \nabla_0\Sigma = \partial_0 \Sigma 
 + i \left(\frac{M M^\dagger}{2p_F}\right)\Sigma
 - i \Sigma\left(\frac{ M^\dagger M}{2p_F}\right) .
\ee
$X_L$ and $X_R$ contribute to the vacuum energy at $O(M^4)$
\be
\label{E_m4}
\Delta {\cal E} = \frac{f_\pi^2}{8p_F^2} 
 {\rm Tr}\left[(MM^\dagger)(M^\dagger M)-(MM^\dagger)^2\right].
\ee
This result can also be derived directly in the microscopic 
theory \cite{Bedaque:2001je}. The corresponding diagrams are
shown in Fig.~\ref{fig_m4}. We also note that equation
(\ref{E_m4}) has the expected scaling behavior ${\cal E}
\sim f_\pi^2\Delta^2 [M^2/(p_F\Delta)]^2$.

\begin{figure}[t]
\epsfxsize=12.5cm
\epsffile{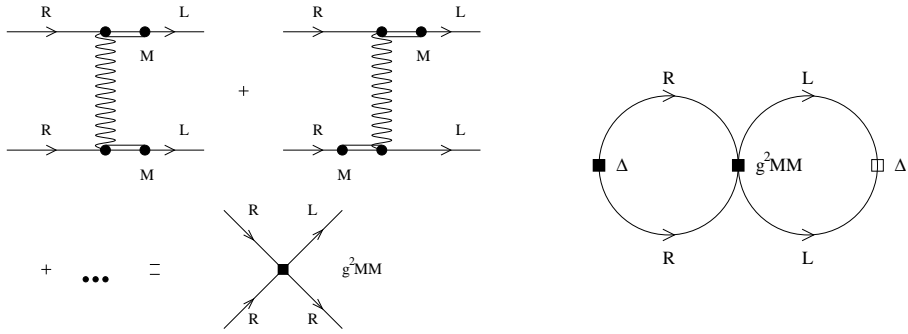}
\vspace*{0.5cm}
\caption{\label{fig_4f}
Fig. a) shows the effective chirality violating four-fermion
vertex in the high density effective theory. Fig. b) shows
the corresponding contribution to the vacuum energy. As 
explained in the text, there are additional vertices which
involve $MM^\dagger$, but they do not contribute to the 
masses of Goldstone modes.}
\end{figure}

 $O(M^2)$ terms in the vacuum energy are generated by terms 
in the high density effective theory that are higher order 
in the $1/p_F$ expansion. We recently argued that these 
terms can be determined by computing chirality violating 
quark-quark scattering amplitudes for fermions in the 
vicinity of the Fermi surface \cite{Schafer:2001za}. 
Feynman diagrams for $q_L+q_L\to q_R+q_R$ are shown in 
Fig.~\ref{fig_4f}a. To leading order in the $1/p_F$
expansion the chirality violating scattering amplitudes
are independent of the scattering angle and can be 
represented as local four-fermion operators
\be
\label{hdet_m}
 {\cal L} = \frac{g^2}{8p_F^4}
 \left( ({\psi^A_L}^\dagger C{\psi^B_L}^\dagger)
        (\psi^C_R C \psi^D_R) \Gamma^{ABCD} +
        ({\psi^A_L}^\dagger \psi^B_L) 
        ({\psi^C_R}^\dagger \psi^D_R) \tilde{\Gamma}^{ACBD} \right).
\ee
There are two additional terms with $(L\leftrightarrow R)$ and
$(M\leftrightarrow M^\dagger)$. We have introduced the CFL eigenstates 
$\psi^A$ defined by $\psi^a_i=\psi^A (\lambda^A)_{ai}/\sqrt{2}$, $A=0,
\ldots,8$. The tensors $\Gamma$ is defined by
\bea 
 \Gamma^{ABCD} &=& \frac{1}{8}\Big\{ {\rm Tr} \left[ 
    \lambda^A M(\lambda^D)^T \lambda^B M (\lambda^C)^T\right] 
  \nonumber \\
 & & \hspace{1cm}\mbox{}
   -\frac{1}{3} {\rm Tr} \left[
    \lambda^A M(\lambda^D)^T \right]
    {\rm Tr} \left[
    \lambda^B M (\lambda^C)^T\right] \Big\}.
\eea
The second tensor $\tilde{\Gamma}$ involves both $M$ and $M^\dagger$
and only contributes to terms of the form ${\rm Tr}[MM^\dagger]$ 
in the vacuum energy. These terms do not contain the chiral
field $\Sigma$ and therefore do not contribute to the masses
of Goldstone modes. We can now compute the shift in the vacuum 
energy due to the effective vertex (\ref{hdet_m}). The leading 
contribution comes from the two-loop diagram shown in 
Fig.~\ref{fig_4f}b. We find
\be
\label{E_MM}
\Delta {\cal E} = -\frac{3\Delta^2}{4\pi^2} 
 \left\{  \Big( {\rm Tr}[M]\Big)^2 -{\rm Tr}\Big[ M^2\Big]
   \right\}
 + \Big(M\leftrightarrow M^\dagger \Big).
\ee
Using this result we can determines the coefficients $A_{1,2,3}$ 
in the CFL chiral theory. We obtain
\be
 A_1= -A_2 = \frac{3\Delta^2}{4\pi^2}, 
\hspace{1cm} A_3 = 0,
\ee
which agrees with the result of Son and Stephanov \cite{Son:1999cm}.
We also note that ${\cal E}\sim f_\pi^2\Delta^2 (\Delta/p_F)$
$[M^2/(p_F\Delta)]$ which shows that the coefficients $A_i$ 
are suppressed by $(\Delta/p_F)$. The effective lagrangian
(\ref{hdet}) and (\ref{hdet_m}) can also be used to compute 
higher order terms in $M$. The dominant $O(M^4)$ term is the
effective chemical potential term equ.~(\ref{E_m4}). Other
$O(M^4)$ terms are suppressed by additional powers of $(\Delta/
p_F)$.

\subsection{Kaon Condensation}
\label{sec_kcond}

 Using the results discussed in the previous section we
can compute the masses of Goldstone bosons in the CFL phase.
In section 3.1 we argued that the expansion parameter 
in the chiral expansion of the Goldstone boson masses is $\delta=
m^2/(p_F\Delta)$. The first term in this expansion comes from the 
$O(M^2)$ term in (\ref{l_cheft}), but the coefficients $A$ contain 
the additional small parameter $\epsilon=(\Delta/p_F)$. In a combined 
expansion in $\delta$ and $\epsilon$ the $O(\epsilon\delta)$ mass 
term and the $O(\delta^2)$ chemical potential term appear at 
the same order. At this order, the masses of the flavored 
Goldstone bosons are
\bea 
\label{mgb}
 m_{\pi^\pm} &=&  \mp\frac{m_d^2-m_u^2}{2p_F} +
         \left[\frac{4A}{f_\pi^2}(m_u+m_d)m_s\right]^{1/2},\nonumber \\
 m_{K_\pm}   &=&  \mp \frac{m_s^2-m_u^2}{2p_F} + 
         \left[\frac{4A}{f_\pi^2}m_d (m_u+m_s)\right]^{1/2}, \\
 m_{K^0,\bar{K}^0} &=&  \mp \frac{m_s^2-m_d^2}{2p_F} + 
         \left[\frac{4A}{f_\pi^2}m_u (m_d+m_s)\right]^{1/2}.\nonumber
\eea
We observe that the pion masses are not strongly affected 
by the effective chemical potential but the masses of the 
$K^+$ and $K^0$ are substantially lowered while the $K^-$ 
and $\bar{K}^0$ are pushed up. As a result the $K^+$ and 
$K^0$ meson become massless if $m_s\sim m_{u,d}^{1/3}\Delta^{2/3}$.
For larger values of $m_s$ the kaon modes are unstable, signaling 
the formation of a kaon condensate. 

 Once kaon condensation occurs the ground state is reorganized.
For simplicity, we consider the case of exact isospin symmetry
$m_u=m_d\equiv m$. Kaon condensation can be studied using an 
ansatz of the form $\Sigma = \exp(i\alpha\lambda_4)$. The 
vacuum energy is 
\be 
\label{k0+_V}
 V(\alpha) = -f_\pi^2 \left( \frac{1}{2}\left(\frac{m_s^2-m^2}{2p_F}
   \right)^2\sin(\alpha)^2 + (m_{K}^0)^2(\cos(\alpha)-1)
   \right),
\ee
where $(m_K^0)^2= (4A/f_\pi^2)m_{u,d} (m_{u,d}+m_s)$ is the $O(M^2)$ 
kaon mass in the limit of exact isospin symmetry. Minimizing the vacuum 
energy we obtain $\alpha=0$ if $m_s^2/(2p_F)<m_K^0$ and $\cos(\alpha)
=(m_K^0)^2/\mu_s^2$ with $\mu_s=m_s^2/(2p_F)$ if $\mu_s
>m_K^0$. We observe that the vacuum energy is independent of 
$\theta_1,\theta_2,\phi$. The hypercharge density is given by
\be 
n_Y = f_\pi^2 \mu_s \left( 1 -\frac{(m_K^0)^4}{\mu_s^4}\right).
\ee
We observe that within the range of validity of the effective 
theory, $\mu_s<\Delta$, the hypercharge density satisfies 
$n_Y<\Delta p_F^2/(2\pi^2)$. This means that the number of condensed 
kaons is bounded by the number of particles contained within a 
strip of width $\Delta$ around the Fermi surface. It also 
implies that near the unlocking transition, $\mu_s\sim
\Delta$, the CFL state is significantly modified. In this
regime, of course, we can no longer rely on the effective 
theory and a more microscopic calculation is necessary. 

  Let us summarize what we have learned about the effects 
of a non-zero strange quark mass $m_s$. The effect of $m_s$ 
is controlled by the parameter $m_s^2/(p_F\Delta)$. If 
$m_s^2/(p_F\Delta)\sim 1$, color-flavor-locking breaks down 
and a transition to a less symmetric phase will occur.
In the regime $m_s^2/(p_F\Delta)<1$ the phase structure can 
be established using the effective chiral theory of the CFL 
phase and dimensional analysis. We have argued that there 
is a new small scale $m_s^2/(p_F\Delta)\sim (\sqrt{m_{u,d}
m_s}/p_F)\ll 1$ which corresponds to the onset of kaon 
condensation. If perturbative QCD is reliable we can be 
more quantitative. To leading order in $g$, the critical 
strange quark mass for kaon condensation is 
\be
\left. m_s \right|_{crit}= 3.03\cdot  m_d^{1/3}\Delta^{2/3} .
\ee
This result suggests that for values of the strange quark mass 
and the gap that are relevant to compact stars CFL matter is 
likely to support a kaon condensate.

\section{Conclusion: The many phases of QCD}
\label{sec_sum}

\begin{figure}[t]
\epsfxsize=12cm 
\epsffile{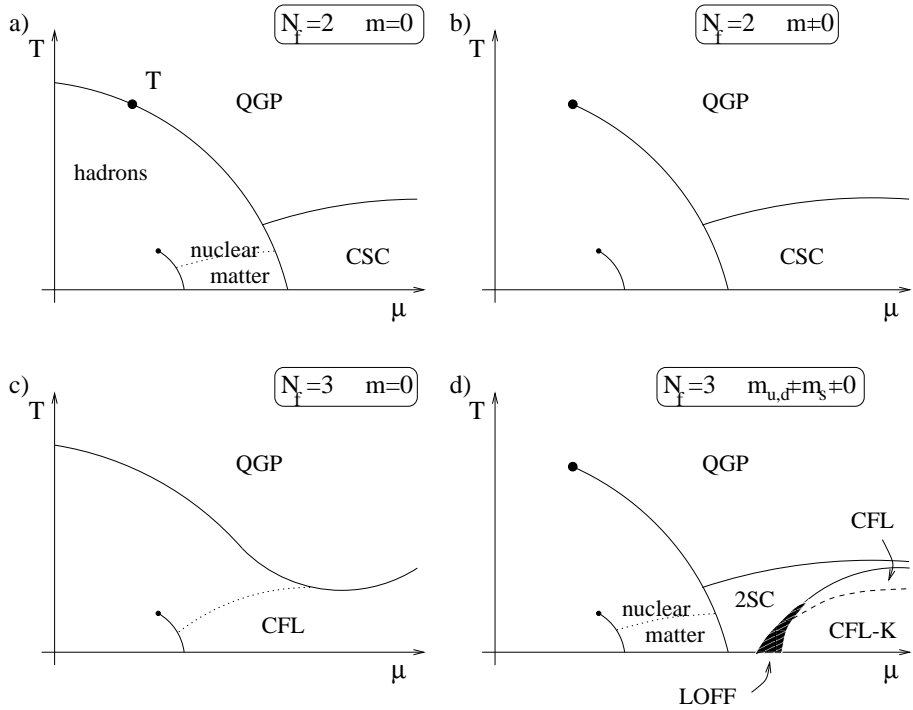}
\vspace*{0.5cm}
\caption{Schematic phase diagram of QCD at finite temperature
and density. The figures a)-d) correspond to different numbers
of massless and massive flavors, see the discussion in the text.  
\label{fig_phase}}
\end{figure}

 We would like to conclude by summarizing some of the things
we have learned about the phase structure of QCD-like theories
at finite temperature and chemical potential. We begin with the 
case of two massless flavors, Fig.~\ref{fig_phase}a. If we
move along the chemical potential axis at temperature $T=0$, there
is a minimum chemical potential required in order to introduce 
baryons into the system. Since nuclear matter is self-bound,
this point is a first order transition. Along the temperature
axis, the line of first order transitions eventually ends in 
a critical point: This is the endpoint of the nuclear liquid-gas
phase transition. If we continue to increase the chemical potential,
we encounter the various phases of nuclear matter at high density. 
Many possibilities have been discussed in the literature, and 
we have nothing to add to this discussion. At even higher 
chemical potential, we encounter the transition to quark matter
and the two flavor quark superconductor. Model calculations
suggest that this transition is first order. In the case of 
two massless flavors, universality arguments suggest, and lattice 
calculations support, the idea that the finite temperature zero 
chemical potential chiral phase transition is second order. In 
this case, the line of first order $\mu\neq 0$ transitions would 
have to meet the $T\neq 0$ transition at a tricritical point 
\cite{BR_98,HJS*_98}. 

  This tricritical point is quite remarkable, because it remains 
a true critical point, even if the quark masses are not zero, 
Fig.~\ref{fig_phase}b. A non-zero quark mass turns the second order 
$T\neq 0$ transition into a smooth crossover, but the first order 
$\mu\neq 0$ transition persists. While it is hard to predict where 
exactly the tricritical point is located in the phase diagram it
may well be possible to settle the question experimentally. 
Heavy ion collisions at relativistic energies produce 
matter under the right conditions and experimental 
signatures of the tricritical point have been suggested
\cite{SRS_98}.

  We have already discussed the phase structure of $N_f=3$
QCD with massless or light degenerate quarks in section
2.2. We emphasized that at $T=0$ the low density,
hadronic, phase and the high density, quark, phase might
be continuously connected. On the other hand, there has
to be a phase transition that separates the color-flavor
locked phase from the $T=\mu=0$ vacuum state. This is
because of the presence of a gauge invariant $U(1)$ order
parameter related to superfluidity that distinguishes the 
two. In the case of $N_f=3$ massless flavors the finite 
temperature phase transition is known to be first order. 
We expect the transition from the superconducting to the 
normal phase at $T\neq 0$ and large $\mu$ to be first 
order, too. This means that there is no tricritical 
point in Fig.~\ref{fig_phase}c. 

  The phase diagram becomes more complicated if we take
into account the effects of a finite strange quark mass,
Fig.~\ref{fig_phase}d. If $m_s^2/(4\mu)\sim\Delta$
there is a phase transition between the $N_f=2+1$
phase with separate pairing in the $ud$ and $ss$ sectors 
and the CFL phase with pairing in both $ud$ and 
$us$ as well as $ds$ sectors. Near this phase
transition we may encounter phases with inhomogeneous
BCS (LOFF) pairing \cite{Alford:2000ze}. Inside the CFL 
phase kaon condensation is a possibility.

  The phase diagram shown in Fig.~\ref{fig_phase}d
should, at best, be considered an educated guess. 
Whether for realistic values of the quark masses there 
is an interlude of the the 2SC phase along the $\mu\neq 0$ 
axis, instead of a direct transition between the CFL phase 
and nuclear matter, cannot be decided on the basis of
currently available calculations. We know that there 
is at least one phase transition, because nuclear matter 
and the color-flavor locked phase are distinguished 
by a gauge invariant $U(1)_s$ order parameter related 
to strangeness. This, of course, is based on our belief 
that ordinary nuclear matter, not strange quark matter 
or hyperonic matter, is the true ground state of 
baryonic matter. 

  Current calculations have also not conclusively
answered the question whether the transition along the
$T\neq 0$ axis is a smooth crossover, as indicated in 
figure \ref{fig_phase}d and favored by some lattice 
calculations, or whether the transition is first order,
as would be the case if $m_s$ is sufficiently small. 
This is clearly an important question in connection 
with the existence of the tricritical point. 

  The challenge ahead of us is to find experimental
observables, both in heavy ion collisions and the 
observation of neutron stars, that will allow us 
to determine the phase diagram of hot and dense 
matter.

\end{document}